# $Y_3Fe_5O_{12}$ Spin Pumping for Quantitative Understanding of Pure Spin Transport and Spin Hall Effect in a Broad Range of Materials (Invited)

Chunhui Du, Hailong Wang, and P. Chris Hammel, Fengyuan Yang

Department of Physics, The Ohio State University, Columbus, OH, 43210, USA

Abstract

Using $Y_3Fe_5O_{12}$ (YIG) thin films grown by our sputtering technique, we study dynamic spin transport in nonmagnetic (NM), ferromagnetic (FM) and antiferromagnetic (AF) materials by ferromagnetic resonance (FMR) spin pumping. From both inverse spin Hall effect (ISHE) and damping enhancement, we determine the spin mixing conductance and spin Hall angle in many metals. Surprisingly, we observe robust spin conduction in AF insulators excited by an adjacent YIG at resonance. This demonstrates that YIG spin pumping is a powerful and versatile tool for understanding spin Hall physics, spin-orbit coupling (SOC), and magnetization dynamics in a broad range of materials.

## I. Introduction

FMR spin pumping is an emerging technique for dynamic injection of a pure spin current from a FM into a NM without an accompanying charge current [1-16], which offers the potential to enable low energy cost, high efficiency spintronics. The performance of these future spin-based applications relies on the efficiency of spin transfer across the FM/NM interfaces [3-16]. YIG has been widely used in microwave applications and is particularly desirable for dynamic spin transport due to its exceptionally low damping [17]. Its insulating nature also allows clean detection of the pure spin current by ISHE in the NM. Here we present our recent results on FMR spin pumping using YIG thin films grown by a sputtering technique that we developed for deposition of high quality epitaxial films of complex materials [12, 18-22]. In particular, we focus on the characterization of structural and magnetic quality of the YIG films, spin pumping from YIG into a number of metals and ISHE detection of the spin currents, determination of interfacial spin mixing conductance ($g_{\uparrow\downarrow}$) and spin Hall angle ($\theta_{SH}$), and spin transport in an AF insulator.

## II. Crystal and Magnetic Structures of $Y_3Fe_5O_{12}$

$Y_3Fe_5O_{12}$ has a cubic crystal structure with space group *Ia-3d* as shown in Fig. 1. The cubic unit cell has a lattice constant of $a = 12.376$ Å and contains 8 formula units (f.u.) with 160 atoms, of which only the $Fe^{3+}$ ions carry magnetic moment (5 $\mu_B$ each). Of the 40 $Fe^{3+}$ ions in a unit cell, 16 are on octahedral sites and 24 are on tetrahedral sites. Each octahedral Fe is connected to 6 tetrahedral Fe and each tetrahedral Fe is connected to 4 octahedral Fe through corning sharing an oxygen, resulting in an intertwining octahedron-tetrahedron network. The magnetic moments of all the octahedral Fe are aligned in parallel. The same is true for all the tetrahedral Fe, but in opposite direction to that of the octahedral Fe, resulting in a ferrimagnetic order. The octahedron-tetrahedron Fe network leads to very low magnetic anisotropy, which in turn leads to exceptionally low damping in YIG. We will discuss later the critical importance of preserving the stoichiometry and ordering of YIG, both in the bulk of the films and at the interfaces, for achieving high-efficiency spin pumping.

## III. Experimental Details

The attractive properties of YIG such as low damping and magnetic softness require stoichiometric samples with a high degree of crystalline perfection and magnetic ordering. Liquid-phase epitaxy (LPE) has been the dominant technique for growing YIG epitaxial films and single crystals in the past few decades [23]. Pulsed laser deposition (PLD) has recently been used to grow epitaxial YIG thin films [24-26]. Using a new off-axis sputtering approach, we deposit YIG thin films on $Gd_3Ga_5O_{12}$ (GGG) substrates in a custom ultrahigh vacuum sputtering system [12, 27]. Our sputtering technique is different from conventional on-axis sputtering [28] and high pressure off-axis sputtering [29]. For on-axis sputtering geometry, the energetic bombardment of the sputtered atoms limits the crystalline quality and ordering of the films. For high pressure off-axis sputtering at ~200 mTorr [29], the frequent scattering by the sputter gas typically results in off-stoichiometry in the films, degrading the film quality of complex materials [22]. Our low-pressure (~10 mTorr) off-axis sputtering technique simultaneous minimizes the bombardment damage and maintains the desired stoichiometry in the films.

We determine the crystalline quality of the YIG films by a Bruker triple-axis x-ray diffractometer (XRD) and measure the surface smoothness by a Bruker atomic force microscope (AFM). FMR absorption and spin pumping measurements are performed using a Bruker electron paramagnetic resonance (EPR) spectrometer in a cavity at a radio-frequency (rf) $f = 9.65$ GHz. We measure the frequency dependencies of the FMR linewidth using a microstrip transmission line at a frequency range between 10 and 20 GHz. Magnetic hysteresis loops of our YIG films are taken by a LakeShore vibrating sample magnetometer (VSM).

## IV. Results and Discussion

### IV(a).  Structural and magnetic quality of the YIG films

Figure 2(a) shows a $2\theta$-$\omega$ XRD scan of a 50-nm YIG films deposited on GGG (111), which exhibits clear Laue oscillations, reflecting the high crystalline quality of the YIG film. The out-of-plane lattice constant of the YIG film is determined to be $c$ = 12.383 Å which is identical to the lattice constant of GGG ($a$ = 12.383 Å) and only 0.06% larger than the bulk value ($a$ = 12.376 Å) of YIG. The full-width-at-half-maximum (FWHM) of a XRD rocking curve is a widely used measure of crystalline quality for epitaxial films. The left inset to Fig. 2(a) give a FWHM of 0.0072° for the first Laue oscillation peak to the left of YIG (444), demonstrating the state-of-the-art crystalline uniformity of the YIG film. The right inset to Fig. 2(a) shows an x-ray reflectometry (XRR) scan of a YIG/Pt bilayer with two periods of oscillations, corresponding to the 34-nm YIG and 4.1-nm Pt layers. A fit to the XRR scan gives a YIG/Pt interfacial roughness of 0.22 nm, indicating the sharpness of the interface. The smooth surface of the YIG films is confirmed by the AFM image in Fig. 2(b) from which we obtain a root-mean-square (rms) roughness of 0.10 nm over an area of 10 μm × 10 μm. Figure 2(c) shows a room temperature in-plane magnetic hysteresis loop for a 20-nm YIG film, which exhibits a very small coercivity ($H_c$) of 0.35 Oe and exceptionally sharp reversal: the magnetic switching is completed within 0.1 Oe. This indicates the high magnetic uniformity of the YIG film. The high crystalline and magnetic uniformity of our YIG films provide the material platform for spin pumping study of a broad range of materials.

### IV(b).  FMR spin pumping of YIG/metal bilayers

Figure 3(a) shows the derivative of a FMR absorption spectrum of a YIG(20 nm) film at an rf power $P_{rf}$ = 0.2 mW in an in-plane field, which gives a peak-to-peak linewidth $\Delta H$ = 7.4 Oe.

Figure 3(b) illustrates the geometry for ISHE detection of spin pumping in a FMR cavity where a DC magnetic field **H** is applied in the $xz$ plane at an angle $\theta_H$ with respect to the film surface and an rf field $h_{rf}$ is applied along the $y$-axis. All samples are approximately 5 mm long and 1 mm wide. An ISHE voltage ($V_{ISHE}$) vs. $H$ plot for a YIG/Pt(5 nm) bilayer at $P_{rf}$ = 200 mW is shown in Fig. 3(c) which exhibits $V_{ISHE}$ = 1.74 mV and antisymmetric dependence on $H$ as expected from the ISHE. $V_{ISHE}$ has a linear $P_{rf}$ dependence [left inset to Fig. 3(c)], indicating that the spin pumping is in the linear regime up to 200 mW. The right inset to Fig. 3(c) shows a sinusoidal behavior of the normalized $V_{ISHE}$ as a function of $\theta_H$, which is expected from spin pumping.

We measure the ISHE in a broad range of metals under the same FMR conditions. Figure 4 shows our results for Cr(5 nm), Fe(10 nm), Co(10 nm), $Ni_{80}Fe_{20}$ [Py(5 nm)], Ni(10 nm), Cu(10 nm), Nb(10 nm), Ag(5 nm), Ta(5 nm), W(5 nm), Pt(5 nm), and Au(5 nm) on YIG [12, 14, 16, 30], which give $V_{ISHE}$ = -5.01 mV, -13.5 μV, -23.5 μV, +70.5 μV, +42.1 μV, +1.06 μV, -666 μV, +1.60 μV, -5.10 mV, -5.26 mV, +3.04 mV, and +72.6 μV, respectively. The sign and magnitude of $V_{ISHE}$ reflects the variation in $\theta_{SH}$, which can be calculated from [5, 8, 10, 11, 14],

$$V_{ISHE} = \frac{-e\theta_{SH}}{\sigma_{NM} t_{NM}} \lambda_{SD} \tanh\left(\frac{t_{NM}}{2\lambda_{SD}}\right) g_{\uparrow\downarrow} f L P \left(\frac{\gamma h_{rf}}{4\pi\alpha f}\right)^2, \tag{1}$$

where $e$ is the electron charge, $\sigma_{NM}$, $t_{NM}$, $\lambda_{SD}$, and $L$ are the conductivity, thickness, spin diffusion length, and sample length of the NM layer, respectively, $h_{rf}$ = 0.25 Oe [14] in our FMR cavity at $P_{rf}$ = 200 mW, $\gamma$ is the gyromagnetic ratio, and $\alpha$ is the Gilbert damping constant of YIG. The factor $P$ = 1.21 [14] arises from the ellipticity of the magnetization precession.

Eq. (1) indicates that calculation of $\theta_{SH}$ relies on accurate measurement of $g_{\uparrow\downarrow}$, which can vary from sample to sample. In literatures, it is uncommon that $V_{ISHE}$ and $g_{\uparrow\downarrow}$ are measured independently. Our thin YIG films with narrow FMR linewidth and low damping allows us to independently determine $g_{\uparrow\downarrow}$ from the damping enhancement [2, 5, 6, 8, 11, 14],

$$g_{\uparrow\downarrow} = \frac{4\pi M_s t_{\text{YIG}}}{g\mu_B}\left(\alpha_{\text{YIG/NM}} - \alpha_{\text{YIG}}\right) \qquad (2)$$

where $g$, $\mu_B$, $M_s$ and $t_{\text{YIG}}$ are the Landé $g$ factor, Bohr magneton, saturation magnetization and thickness of the YIG films, respectively. We obtain the damping constants from the frequency dependencies of the FMR linewidth: $\Delta H = \Delta H_{\text{inh}} + \frac{4\pi\alpha f}{\sqrt{3}\gamma}$ [31], where $\Delta H_{\text{inh}}$ is the inhomogeneous broadening as shown in Fig. 5 for several YIG-based structures. Using least-squares fits to the data in Fig. 5, we obtain the damping constants [$\alpha_{\text{YIG}} = (8.7 \pm 0.6) \times 10^{-4}$ for a bare YIG film] for each structure. Table I lists the values of $g_{\uparrow\downarrow}$ for 13 metals on YIG that we have studied [12, 14, 16, 30], among which the YIG/Pt exhibits the highest $g_{\uparrow\downarrow}$ of $(6.9 \pm 0.6) \times 10^{18}$ m$^{-2}$.

Regarding $\lambda_{\text{SD}}$, since the term $\lambda_{\text{SD}}\tanh(\frac{t_{\text{NM}}}{2\lambda_{\text{SD}}})$ in Eq. (1) is virtually insensitive to the value of $\lambda_{\text{SD}}$ when $\lambda_{\text{SD}} \geq t_{\text{NM}}$ [14], we only need to measure $\lambda_{\text{SD}}$ for materials with short $\lambda_{\text{SD}}$. For those with long $\lambda_{\text{SD}}$, such as Cu, we use values reported in literature. Figure 6(a) plots Pt thickness ($t_{\text{Pt}}$) dependence of the ISHE-induced charge current $I_c = V_{\text{ISHE}}/Rw$ for YIG/Pt bilayers, where $R$ and $w$ are the resistance and width of the YIG/Pt samples. Given that $I_c$ is proportional to the pure spin current pumped into Pt, we obtain $\lambda_{\text{SD}} = 7.3 \pm 0.8$ nm for Pt by fitting to $\frac{V_{\text{ISHE}}}{Rw} \propto \lambda_{\text{SD}}\tanh\left(\frac{t_{\text{Pt}}}{2\lambda_{\text{SD}}}\right)$ [32]. Similarly, we calculate $\lambda_{\text{SD}} = 1.9 \pm 0.2$, $2.1 \pm 0.2$, and $13.3 \pm 2.1$ nm for Ta, W, and Cr, as shown in Figs. 6(b), 6(c), and 6(d), respectively. Using the obtained values of $g_{\uparrow\downarrow}$ and $\lambda_{\text{SD}}$, we calculate $\theta_{\text{SH}}$ for the 13 metals (Table I) [14, 16, 30], which reveal the important role of $d$-electron configuration of transition metals in spin Hall effect (SHE). This is consistent with the prediction of Tanaka $et$ $al.$ [33] who calculated the spin Hall conductivities (SHC) of the $4d$ ($5d$) transition metals by considering the role of the total number of $4d$ ($5d$) and $5s$ ($6s$) electrons. We list in Table I the total number of $d$ and $s$ electrons, $n_{d+s}$, in the conduction bands and plot $\theta_{\text{SH}}$ vs. $n_{d+s}$ for four $3d$ and four $5d$ metals in Fig. 7. The calculated SHC of $5d$ metals [33] are also shown for comparison. Both $3d$ and $5d$ metals show the same systematic behavior in $\theta_{SH}$ which varies significantly both in sign and magnitude as a function of $n_{d+s}$. While the behavior of $\theta_{SH}$ in $4d$ and $5d$ transition metals are understood theoretically [33, 34], theoretical calculations for spin Hall effect in $3d$ metals are needed to explain the observed ISHE in $3d$ metals.

### IV(c). Spin transport in antiferromagnetic insulators

Following the spin pumping study from YIG into metals, we investigate spin transport in YIG/insulator/Pt trilayer systems, where the insulator spacer is either a diamagnet, $SrTiO_3$ (STO), or an AF, NiO. Figure 8(a) shows a semi-log plot of $V_{ISHE}$ as a function of the $SrTiO_3$ thickness ($t_{STO}$) in YIG/STO/Pt(5 nm) trilayers, which exhibit a clear exponential decay with a decay length of $\lambda = 0.19$ nm [13]. The short decay length is comparable to the inter-atomic spacing, which indicates that the Pt conduction electrons tunnel across the STO barrier and exchange couple to the precessing YIG magnetization to acquire spin polarization. This result demonstrates that exchange coupling is the dominant mechanism responsible for spin pumping [2, 13]. More interestingly, when an AF insulator NiO is used, spin currents can propagate over a much longer distance, as shown in Fig. 8(b) for YIG/NiO/Pt trilayers which exhibits a decay length of $\lambda = 9.4$ nm [35]. Strikingly, we observe a significant enhancement in ISHE voltages at small NiO thicknesses $t_{NiO}$: $V_{ISHE}$ increases from 0.604 mV for the YIG/Pt bilayer to 1.30 mV for the YIG/NiO(2 nm)/Pt trilayer. This is in drastic contrast to the suppression of $V_{ISHE}$ by more than two orders of magnitude when a 1-nm $SrTiO_3$ is inserted between YIG and Pt. The surprising enhancement of $V_{ISHE}$, long spin decay length in YIG/NiO/Pt structures point toward the AF nature of NiO as the underlying mechanism for the observed robust spin transport in NiO [35]. Since the range of $t_{NiO}$ covers NiO layers with ordering temperatures [36] both above (large $t_{NiO}$) and below (small $t_{NiO}$) room temperature, this indicates that both AF ordered and AF fluctuating [37] spins in NiO can be excited by exchange coupling to the precessing YIG magnetization and efficiently transporting spin current over a long distance.

### IV(d). Key factors contributing to large ISHE signals

Lastly, we comment below on several important factors that are critical for achieving high spin current and large ISHE signals. (1) The spin current generated in YIG/NM bilayers depends on how out of equilibrium that the YIG magnetization can be excited, which can be described by the precession cone angle [2, 5, 8]. This demands YIG films with low damping, a widely accepted criterion in spin pumping. (2) However, the measured value of $\alpha$ is a "bulk" property of the YIG film, while spin pumping is determined predominantly by the YIG/NM interface because of the exchange mechanism with an effective distance of ~0.2 nm [2, 13]. As an example, the insertion of a 1-nm SrTiO$_3$ layer between YIG and Pt reduces $V_{ISHE}$ by a factor of 256 [Fig. 8(a)]. This suggests that a large precession cone angle persisting close to the interface with Pt is critical for high-efficiency spin transfer. Thus, the YIG film needs to maintain correct stoichiometry, high crystalline quality, and uniform magnetic ordering from inside to the top atomic layer of the YIG film. Post-deposition treatments, such as polishing, etching, or sometimes annealing, could jeopardize the YIG surface. (3) Spin mixing conductance is a phenomenological parameter that describes the quality of the YIG/NM interface in conducting spin currents. The values of $g_{\uparrow\downarrow}$ vary significantly for the same YIG/Pt structure made by different techniques and research groups due to variation of the interface quality [6, 10, 12]. Since characterizing the chemical, structural, and magnetic uniformity of the YIG surface is rather challenging, ISHE voltage is a good quantity for comparing the spin pumping efficiency and YIG/Pt interface quality. After all, $V_{ISHE}$ is a direct measure of the pure spin currents pumped into Pt. (4) We find that oxygen content during the YIG growth is critical for its spin pumping performance. Our best YIG thin films for spin pumping can only be grown within a narrow window of the oxygen partial pressure, outside which, the ISHE signals degrade dramatically.  Some of these points are supported by experimental evidence and some are our speculations.  More thorough characterizations of the YIG/NM interfaces are needed to understand the nature of spin transfer from dynamically excited YIG to metals.

## V.  Conclusion

Epitaxial YIG thin films provide a material platform for high-sensitivity investigation of dynamic spin transport in a broad range of nonmagnetic, ferromagnetic, and antiferromagnetic materials, be they conducting or insulating.  The phenomena uncovered in these materials and structures provide guidelines for potential spintronics applications and enable further understanding of fundamental interactions such as spin-orbit coupling and magnetic excitations.

## VI. Acknowledgements

This work was primarily supported by the U.S. Department of Energy (DOE), Office of Science, Basic Energy Sciences, under Award No. DE-FG02-03ER46054 (FMR and spin pumping characterization) and No. DE-SC0001304 (sample synthesis and magnetic characterization).  This work was supported in part by the Center for Emergent Materials, an NSF-funded MRSEC under award No. DMR-1420451 (structural characterization).  Partial support was provided by Lake Shore Cryogenics Inc. and the NanoSystems Laboratory at the Ohio State University.

**Table I.** Interfacial spin mixing conductance $g_{\uparrow\downarrow}$, spin diffusion length $\lambda_{\mathrm{SD}}$, spin Hall angle $\theta_{\mathrm{SH}}$, and total number of $d$ and $s$ electrons in conduction bands, $n_{d+s}$, for metals and alloys studied by our YIG spin pumping.

| Structure | $g_{\uparrow\downarrow}(\mathrm{m}^{-2})$ | $\lambda_{\mathrm{SD}}$ (nm) | $\theta_{\mathrm{SH}}$ | $n_{d+s}$ |
|---|---|---|---|---|
| YIG/**Ti** | $(3.5 \pm 0.3) \times 10^{18}$ | $\sim$13.3 | $-(3.6 \pm 0.4) \times 10^{-4}$ | 4 |
| YIG/**V** | $(3.1 \pm 0.3) \times 10^{18}$ | $\sim$13.3 | $-(1.0 \pm 0.1) \times 10^{-2}$ | 5 |
| YIG/**Cr** | $(8.3 \pm 0.7) \times 10^{17}$ | 13.3 | $-(5.1 \pm 0.5) \times 10^{-2}$ | 6 |
| YIG/**Mn** | $(4.5 \pm 0.4) \times 10^{18}$ | $\sim$13.3 | $-(1.9 \pm 0.1) \times 10^{-3}$ | 7 |
| YIG/**FeMn** | $(4.9 \pm 0.4) \times 10^{18}$ | 3.8 [38] | $-(7.4 \pm 0.8) \times 10^{-5}$ | 7.5 |
| YIG/Cu/**Py** | $(6.3 \pm 0.5) \times 10^{18}$ | 1.7 [30] | $(2.0 \pm 0.5) \times 10^{-2}$ | 9.6 |
| YIG/Cu/**Ni** | $(2.0 \pm 0.2) \times 10^{18}$ | 3.2 [16] | $(4.9 \pm 0.5) \times 10^{-2}$ | 10 |
| YIG/**Cu** | $(1.6 \pm 0.1) \times 10^{18}$ | 500 [39] | $(3.2 \pm 0.3) \times 10^{-3}$ | 11 |
| YIG/**Ag** | $(5.2 \pm 0.5) \times 10^{17}$ | 700 [40] | $(6.8 \pm 0.7) \times 10^{-3}$ | 11 |
| YIG/**Ta** | $(5.4 \pm 0.5) \times 10^{18}$ | 1.9 | $-(6.9 \pm 0.6) \times 10^{-2}$ | 5 |
| YIG/**W** | $(4.5 \pm 0.4) \times 10^{18}$ | 2.1 | $-(1.4 \pm 0.1) \times 10^{-1}$ | 6 |
| YIG/**Pt** | $(6.9 \pm 0.6) \times 10^{18}$ | 7.3 | $(1.0 \pm 0.1) \times 10^{-1}$ | 10 |
| YIG/**Au** | $(2.7 \pm 0.2) \times 10^{18}$ | 60 [39] | $(8.4 \pm 0.7) \times 10^{-2}$ | 11 |

**Figure captions:**

**Figure 1.** Schematic of the cubic unit cell of YIG garnet structure.

**Figure 2.** (a) Semi-log $2\theta$-$\omega$ XRD scan of a 50-nm YIG film grown on a GGG (111) substrate, in which the clear satellite peaks are Laue oscillations. Left inset: XRD rocking curve of the first satellite peak on the left of YIG (444) peak showing a FWHM of 0.0072°. Right inset: X-ray reflectometry spectrum (red) of a YIG(34 nm)/Pt(4.1) bilayer on GGG, where the blue curve is a fit by Bruker Leptos. (b) AFM image of a YIG film with a roughness of 0.10 nm over an area of 10 μm × 10 μm. (c) A room temperature in-plane magnetic hysteresis loop of a 20-nm YIG film which gives a small coercivity of 0.35 Oe and very sharp magnetic reversal (switching completed within 0.1 Oe).

**Figure 3.** (a) Room-temperature derivative of a FMR absorption spectrum of a YIG film with an in-plane field at $P_{rf} = 0.2$ mW, which gives a peak-to-peak linewidth of 7.4 Oe. (b) Schematic of experimental setup for ISHE measurements. (c) $V_{ISHE}$ vs. $H$ spectra at $\theta_H = 90°$ for a YIG(20 nm)/Pt(5 nm) bilayer. Left inset: rf power dependence of $V_{ISHE}$ for the YIG/Pt bilayer, where the blue line is a least-squares fit to the data. Right inset: angular dependence of the normalized $V_{ISHE}$ for the YIG/Pt bilayer, where the red curve is a $\sin\theta_H$ fit.

**Figure 4.** $V_{ISHE}$ vs. $H - H_{res}$ spectra of (a) YIG/Cr(5 nm), (b) YIG/Fe(10 nm), (c) YIG/Co(10 nm), (d) YIG/Py(5 nm), (e) YIG/Ni(10 nm), (f) YIG/Cu(10 nm), (g) YIG/Nb(10 nm), (h)YIG/Ag(5 nm), (i) YIG/Ta(5 nm), (j) YIG/W(5 nm), (k) YIG/Pt(5 nm), and (l) YIG/Au (5 nm) bilayers at $\theta_H = 90°$(red) using $P_{rf} = 200$mW.

**Figure 5.** (a) Frequency dependencies of FMR linewidth of a bare YIG film, YIG/Cr, YIG/Cu, YIG/Au, YIG/W, YIG/Ta, YIG/Pt bilayers, and a YIG/Cu/Py trilayer.

**Figure 6.** ISHE-induced charge current ($V_{ISHE}/R$) normalized by sample width $w$ of (a) YIG/Pt,

(b) YIG/Ta, (c) YIG/W, and (d) YIG/Cr bilayers as a function of the metal layer thicknesses, from which the spin diffusion length of $\lambda_{SD}$ = 7.3 $\pm$ 0.8, 1.9 $\pm$ 0.2, 2.1 $\pm$ 0.2, and 13.3 $\pm$ 2.1 nm, respectively, are obtained.

**Figure 7.** The obtained spin Hall angles ($\theta_{SH}$) as a function of the total number of $d$ and $s$ electrons, $n_{d+s}$, for selected 3d and 5d metals. The theoretical calculations of spin Hall conductivity (SHC) for 5d metals by Tanaka et al. [ref] are also shown (open green circles).

**Figure 8.** (a) Semi-log plot of $V_{ISHE}$ as a function of the SrTiO$_3$ barrier thickness for YIG/SrTiO$_3$/Pt(5 nm) trilayers. (b) Semi-log plot of the NiO thickness dependencies of the ISHE voltages for YIG(20 nm)/NiO($t_{NiO}$)/Pt(5 nm) trilayers. Inset: $V_{ISHE}$ as a function of NiO thickness from 0 to 10 nm.

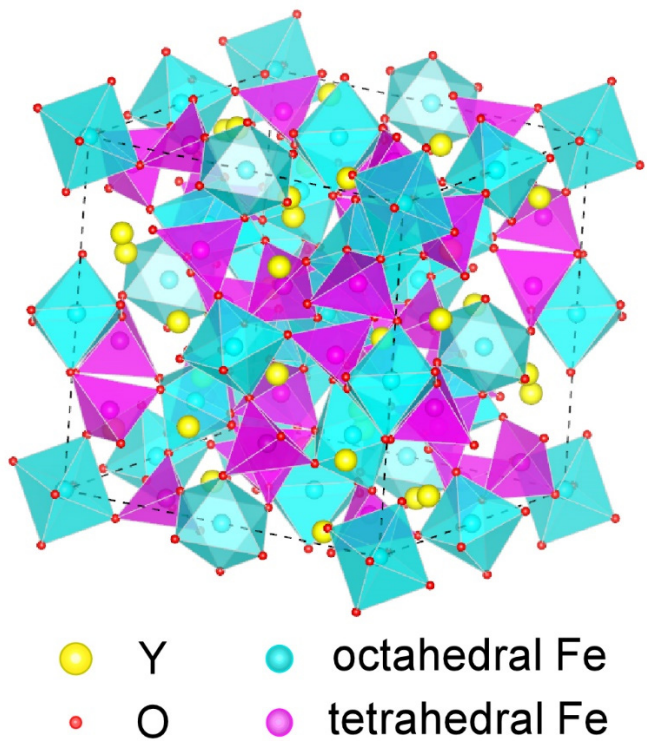

**Figure 1**.

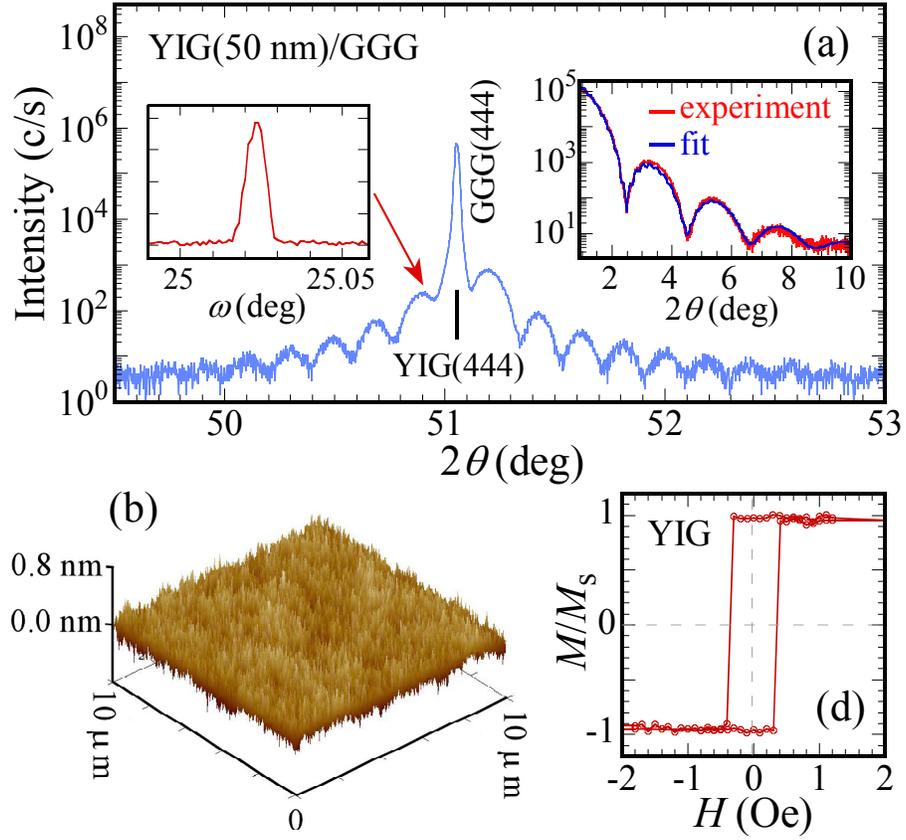

**Figure 2.**

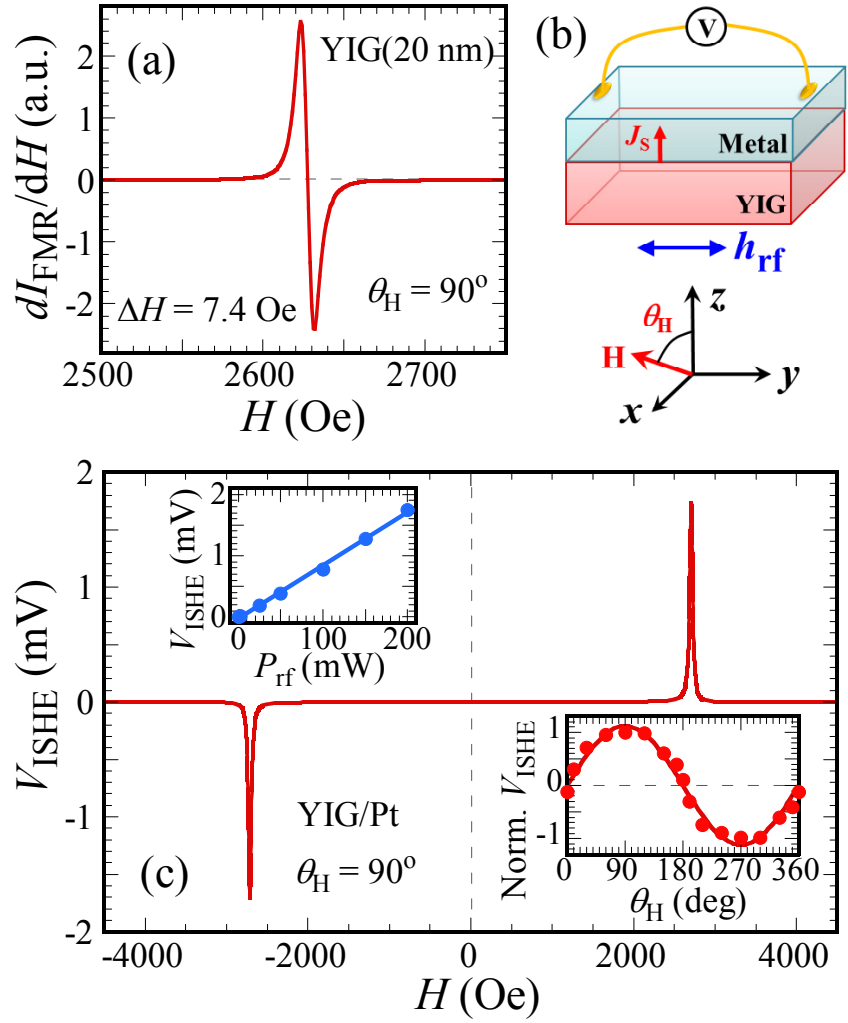

Figure 3

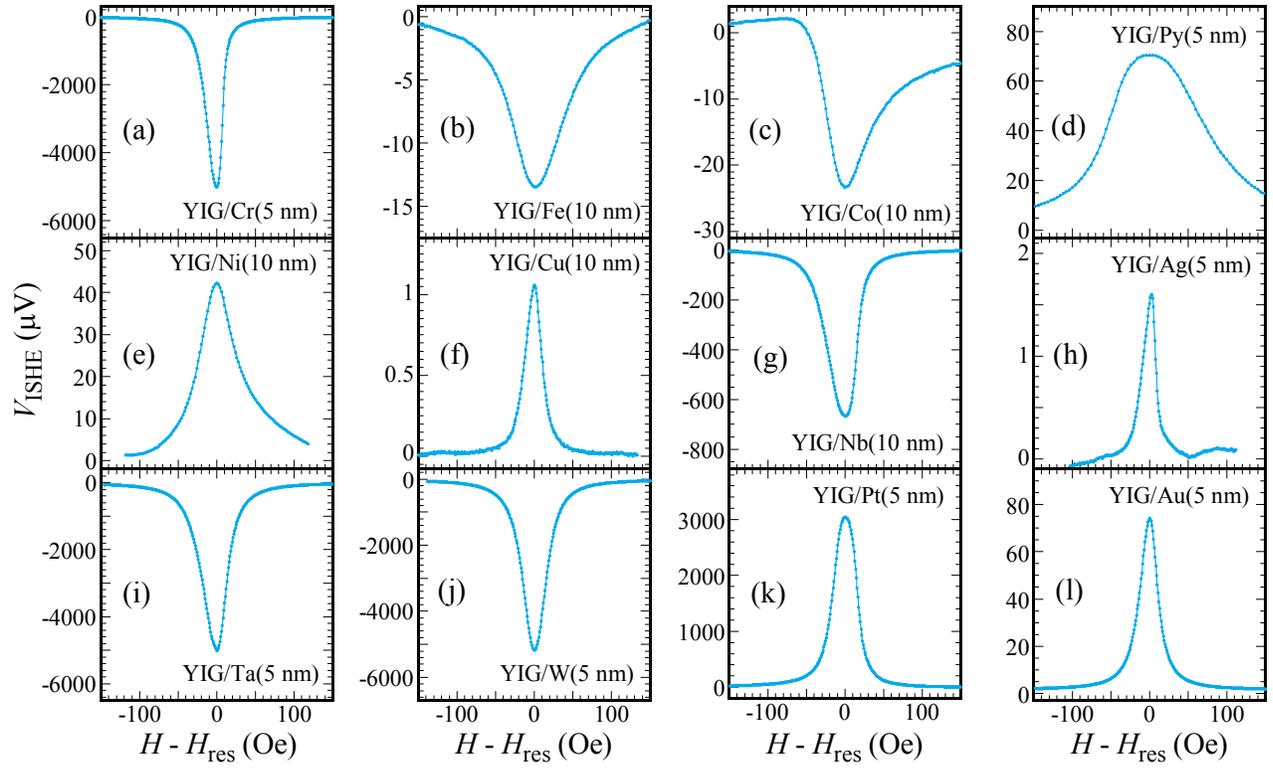

**Figure 4.**

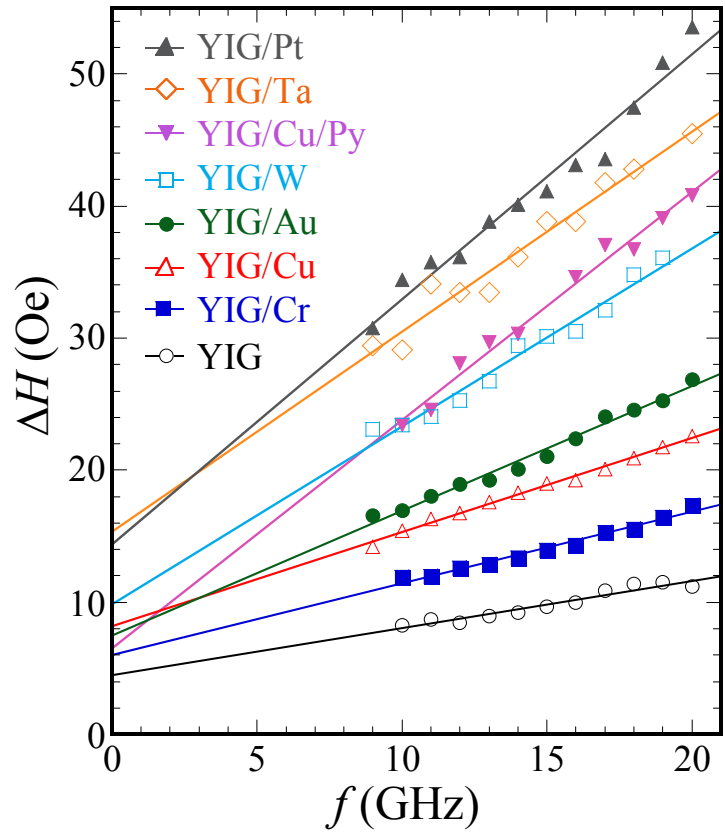

**Figure 5.**

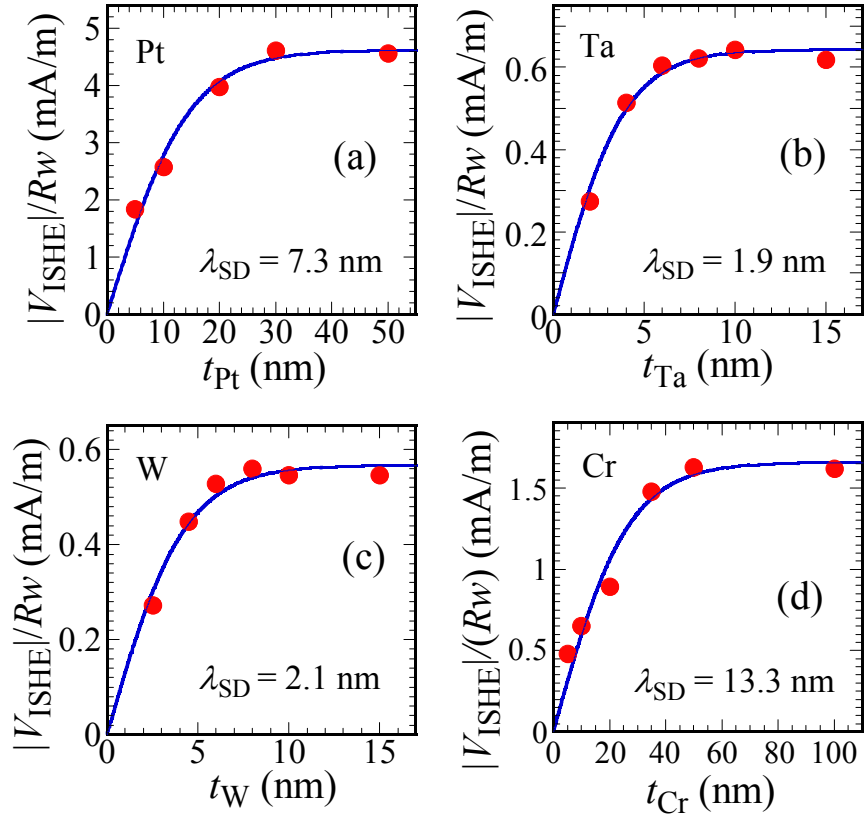

Figure 6.

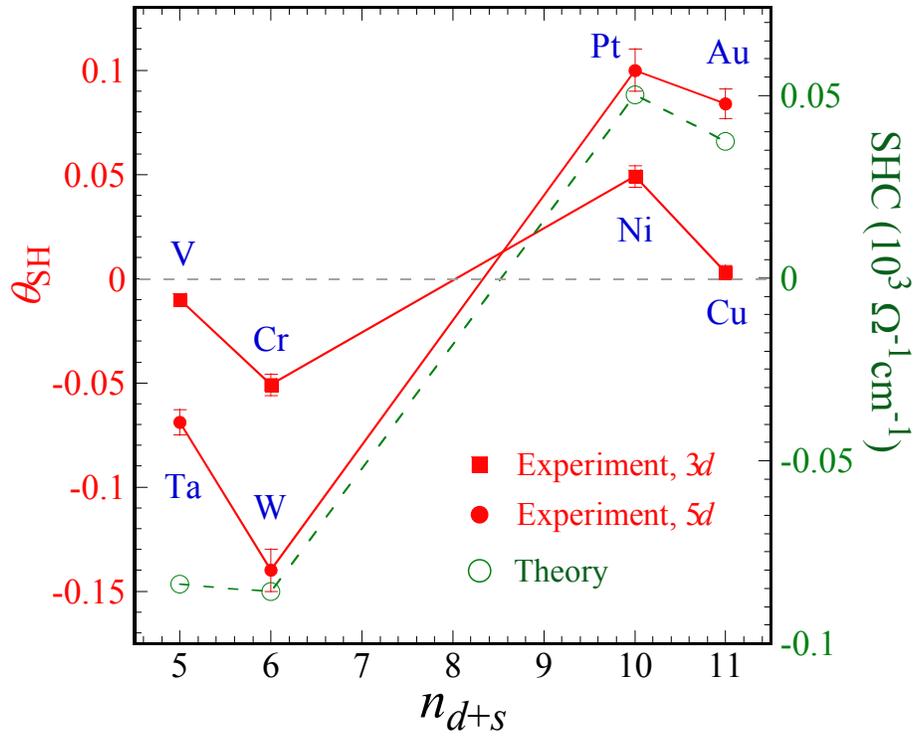

Figure 7

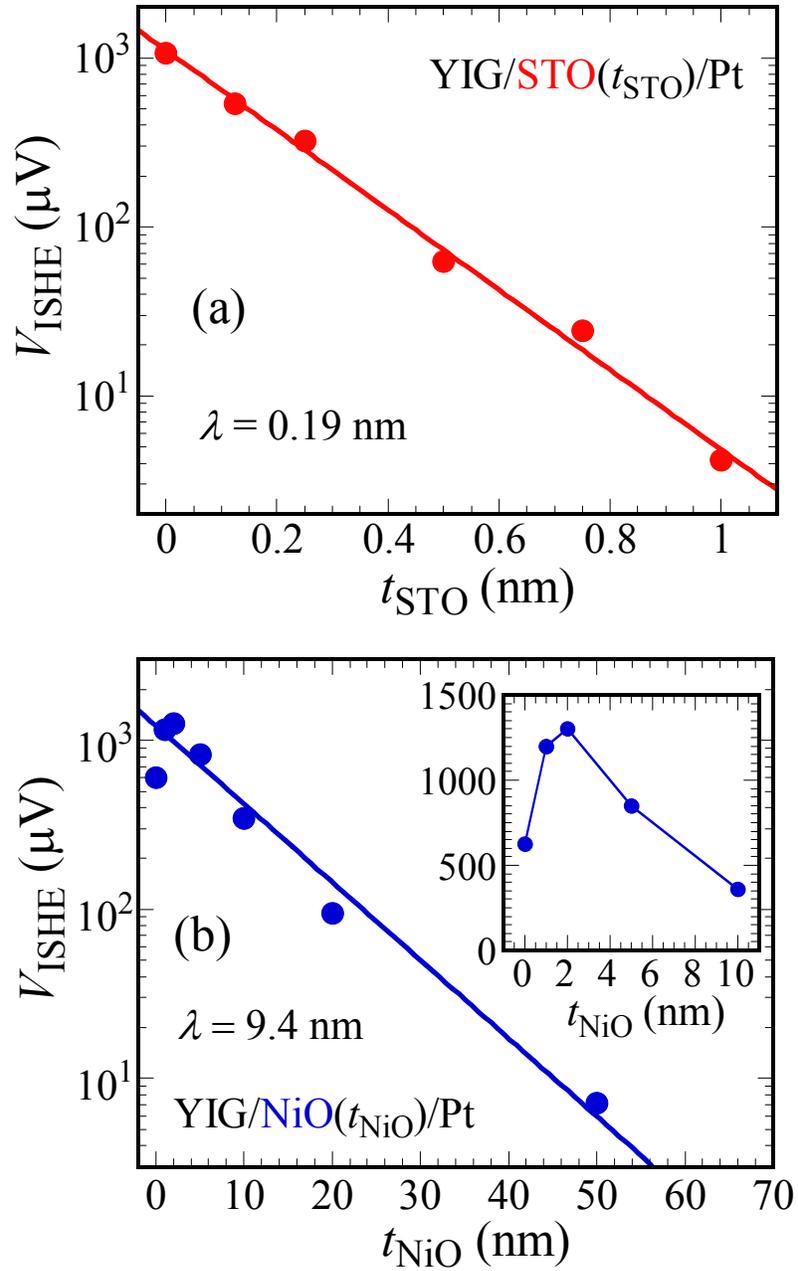